\documentclass[11pt,twoside]{JHEP3}
\preprint{Imperial/TP/09/DJW/01}
\usepackage{graphicx}
\usepackage{amsmath,bbm,amssymb}

\title{Quantum kink and its excitations}
\author{Arttu Rajantie, David J Weir \\ Theoretical Physics, Blackett Laboratory, Imperial College London SW7 2AZ, UK \\ E-mail: \email{a.rajantie@imperial.ac.uk}, \email{david.weir03@imperial.ac.uk}}
\abstract{We show how detailed properties of a kink in quantum field theory can be extracted from field correlation functions. This makes it possible to study quantum kinks in a fully non-perturbative way using Monte Carlo simulations. We demonstrate this by calculating the kink mass as well as the spectrum and approximate wave functions of its excitations. This way of measuring the kink mass has clear advantages over the existing approaches based on creation and annihilation operators or the kink free energy. Our methods are straightforward to generalise to more realistic theories and other defect types. }

\begin{document}

\section{Introduction}

Kinks, domain walls and other topological defects are of interest across physics. The existence of magnetic monopoles, for example, is a general prediction of grand unified theories~\cite{vshel}. Cosmic strings~\cite{Kibble:1976sj} may give an observable contribution to the cosmic microwave background radiation~\cite{Bevis:2007gh} or gravitational wave background~\cite{Siemens:2006yp}. In string theory, D-branes are analogous to topological defects.

There has been widespread interest in the excitation spectra of topological defects, although mostly as a calculational tool for finding perturbative corrections to the mass of the defect~\cite{Dashen:1974cj,Rajamaran1987,Kiselev:1988gf,Kiselev:1998gg,Bessa:2004pu,Mussardo:2005dx,Pawellek:2008st}. However, excitations themselves are of physical interest~\cite{Goldstone:1974gf,Christ:1975wt,Jackiw:1977yn}. Bound states of point-like defects correspond to new particle species, and in the case of extended defects they would be particles propagating on the brane. For instance, in braneworld models all the standard model particles can be thought of as states localised on a domain wall~\cite{Dvali:2000hr}. Finding stable particles localised on defects in fully non-perturbative approaches would therefore correspond to completely new particle species; indeed, for strongly-coupled theories there is no a priori reason to expect that the perturbative spectrum is accurate. The bound states of cosmic strings have been previously studied as a potential way of stabilising string loops~\cite{Goodband:1995nr,Goodband:1995rt,Vachaspati:1992mk}. In some sense, reconstructing the excitation spectrum is the opposite approach to that taken classically in Vachaspati's paper on the reconstruction of a field theory~\cite{Vachaspati:2003vk}.

Domain walls, vortices and strings are also of interest in condensed matter systems. Charge carriers lying on domain walls have been posited as a conduction mechanism~\cite{Nayak1997}.  Interfacial phenomena such as the wetting of a surface can be viewed as the motion of a domain wall in a curved spacetime~\cite{Billo:2006zg,Jasnow1986}. Kinks are also studied in integrable systems~\cite{Lepori:2008et}, and in non-Hermitian field theories~\cite{deSouzaDutra:2007da}.

In this paper we investigate field correlation functions in the presence of a kink, and in this sense our approach bears some similarity with Refs.~\cite{Goldstone:1974gf,Christ:1975wt,Jackiw:1977yn}. We show that the correlators can be used to measure non-perturbatively the mass of the kink and its excitation spectrum. Furthermore, we obtain approximate wavefunctions for the excitations based on Monte Carlo data. Previously these have only been calculated at the linear level, although there has been some recent progress in accounting for interactions using renormalisation group techniques~\cite{Alexandre:2007ci}.

Previous work has measured the kink mass with `Kadanoff's non-local operator'~\cite{Ciria:1993yx}; but more generally, the mass of topological defects has usually been calculated on the lattice by finding the free energy of the defect~\cite{Bursa:2005yv,HijarSutmann:2008,Kajantie:1990bu,Rajantie:2005hi}. This is taken as the difference in free energy between the system with the defect present, and the same system but without the defect. The defect is often created using twisted boundary conditions, so one is effectively measuring the response to that twist~\cite{Groeneveld:1980tt}. Since the partition function cannot be determined by Monte Carlo simulations, one must measure derivatives of the mass and then perform finite differencing from a point where the mass is known. Typically, this is at a phase transition which is where the errors in Monte Carlo simulations are at their largest.

The method of calculating the kink mass we propose requires study only of a topologically nontrivial sector, and only at one parameter choice. This gives a check on errors which may be present in the `twist' method. It may also serve to reduce the amount of calculation required since it only requires Monte Carlo simulations to be carried out for the desired parameters.

Our analytical discussion of correlation functions in the presence of topological defects takes place in Section~\ref{sec:correlationfunctions}. Section~\ref{sec:spectralexpansion} outlines methods which rely on the spectral expansion. We describe how to calculate the mass in Section~\ref{sec:twopointfunction}, while in Section~\ref{sec:fourpointfunction} we determine the particle spectrum of $\lambda \phi^4$ theory and the approximate wavefunctions in the presence of a kink. Monte Carlo simulations and other numerical work based on these ideas are discussed in Section~\ref{sec:numerics}.

We believe that these techniques can be generalised to more sophisticated defects such as cosmic strings and monopoles; the extension to domain walls is trivial. Eventually, similar techniques may be used to study the defects present in supersymmetric theories such as the Wess-Zumino Model or SQCD and hence as a test of lattice supersymmetry~\cite{Giedt:2007hz}.

\subsection{The $\lambda\phi^4$ model}
\label{sec:introduction}

Kinks, or their higher dimensional counterparts domain walls, can occur where the vacuum manifold after spontaneous symmetry breaking is not simply connected, and the field can take on a different vacuum expectation value in different parts of the space. Domain walls could have formed in the early universe due to random fluctuations leading to different patches of space occupying different, disconnected vacua.

In this paper, we focus on kinks in the $1+1$-dimensional $\lambda\phi^4$ model, with the Lagrangian
\begin{equation}
 \mathcal{L} = \frac{1}{2}\left(\partial_\mu \phi\right)^2 - V(\phi); \quad V(\phi) = - \frac{m^2}{2!}\phi^2 + \frac{\lambda}{4!}\phi^4
\end{equation}
with $m^2>0$ in the classical broken phase. The vacuum is then $\phi_0 = \pm m \sqrt{6/\lambda}$. In $(1+1)$ dimensions, the dimensionless parameter that appears in perturbative results will be $\lambda/m^2$. 

A kink configuration interpolates between the two vacua. We can create a topological kink by requiring that $\phi \to \phi_0$ as $x\to +\infty$ and $\phi \to -\phi_0$ as $x\to -\infty$, with one choice of sign being termed a `kink' and the other an `antikink'. The kink solution is obtained via the Bogomol'nyi equation for the kink~\cite{Rajamaran1987},
\begin{equation}
\label{eq:bogomolnyi}
 \frac{\partial \phi}{\partial x} = \pm \sqrt{2V(\phi)}.
\end{equation}
The static classical kink solution $\phi_{\mathrm{k}}$ is a minimum-energy configuration satisfying  (\ref{eq:bogomolnyi}) that interpolates between the two vacua in the broken phase. In the frame of the kink,
\begin{equation}
\label{eq:kinksolution}
 \phi_{\mathrm{k}}(x,t) = m \sqrt{\frac{6}{\lambda}} \tanh \left(\frac{mx}{\sqrt{2}}\right).
\end{equation}
The classical mass $M_{\mathrm{cl}}$ is, from integrating the energy density,
\begin{equation}
\label{eq:classicalmass}
M_{\mathrm{cl}} = 4\sqrt{2}\frac{m^3}{\lambda}.
\end{equation}

\subsection{Bound and scattering states}
\label{sec:boundstates}
We wish to consider the excitation spectrum in the presence of a kink. Expanding about the classical kink solution, $\phi = \phi_{\mathrm{k}} + \hat{\phi}$ and eliminating terms using the equation of motion for the classical kink yields~\cite{Alexandre:2007ci}

\begin{equation}
\label{eq:lagrangiankinkbg}
\mathcal{L} = \frac{1}{2}\left(\partial_\mu \hat{\phi}\right)^2 - \left[ \frac{3m^2}{2}\tanh^2 \left(\frac{mx}{\sqrt{2}}\right) - \frac{m^2}{2} \right]\hat{\phi}^2 + \mathcal{L}_\mathrm{int},
\end{equation}
leading to the form of the action, after integration by parts~\cite{Rajamaran1987}
\begin{equation}
S = \frac{1}{2}\int d^2 x \: \hat{\phi}(x)\left[-\partial^2 + m^2 - 3 m^2 \tanh^2\left(\frac{mx}{\sqrt{2}}\right) \right] \hat{\phi}(x) + S_\mathrm{int}
\end{equation}
where $S_\mathrm{int}$ are terms of order $\hat{\phi}^3$ and above. The Green's function $G$ is given by
\begin{equation}
 \left[\partial^2 - m^2 + 3 m^2 \tanh^2\left(\frac{mx}{\sqrt{2}}\right) \right] G(x_1,t_1;x_2,t_2) = \delta(x_1-x_2)\delta(t_1-t_2).
\end{equation}
We can construct this Green's function out of the eigenfunctions for the corresponding Schr\"{o}dinger equation. Time-translation symmetry is not broken by the kink, so a suitable complete set in the time direction is $\exp(iE_n t)$. On a timeslice, we must solve the Schr\"{o}dinger equation with a P\"{o}schl-Teller potential:
\begin{equation}
\label{eq:bschr}
\left[-\frac{\partial^2}{\partial x^2} + 3 m^2 \tanh^2 \frac{mx}{\sqrt{2}} - m^2 \right]\psi_n(x) = E_n \psi_n(x)
\end{equation}
with the eigenfunctions
\begin{eqnarray}
\psi_0 & = & \cosh^{-2} \left(\frac{mx}{\sqrt{2}}\right) \\
\psi_1 & = & \cosh^{-2} \left(\frac{mx}{\sqrt{2}}\right) \sinh \left(\frac{mx}{\sqrt{2}}\right) \label{eq:bspsi} \\  
\psi_{q} & = & \exp\left({\frac{i q mx}{\sqrt{2}}}\right) \left[3\tanh^2\left(\frac{mx}{\sqrt{2}}\right) - 1 - q^2 - 3iq \tanh\left(\frac{mx}{\sqrt{2}}\right)\right] \label{eq:scatpsi}
\end{eqnarray}
and corresponding energy eigenvalues
\begin{eqnarray}
E_0 & = & 0 \label{eq:goldstoneenergy} \\
E_1 & = & m\sqrt{\frac{3}{2}} \\
E_q & = & m\sqrt{\frac{q^2}{2} + 2} \label{eq:scatteringenergy}.
\end{eqnarray}
$E_0$ corresponds to a Goldstone mode that shifts the kink in the $x$-direction; $E_1$ corresponds to a massive localised `bound state' and $E_q$ are a continuum of scattering states labelled by $-\infty < q < \infty$.  We impose antiperiodic boundary conditions on a box of length $L$ with $L\to \infty$. Then, the following expression for the allowed parameters $q$ can be found, based on odd integer multiples of $\pi/L$ plus a phase shift due to the kink
\begin{equation}
\label{eq:allowedant}
 q_{n,\mathrm{ant}}\frac{mL}{\sqrt{2}} = (2n+1)\pi + \delta(q_{n,\mathrm{ant}}); \qquad n = 0, 1, 2, \ldots
\end{equation}
where $\delta(q)$ is the phase shift experienced by particles scattered off the kink~\cite{Rajamaran1987},
\begin{equation}
\label{eq:phaseshift}
 \delta(q_n) = -2\tan^{-1} \frac{3q_n}{2-q_n^2}.
\end{equation}

The allowed values merge into a continuum when $L\gg 1/m$. Thus, in the infinite volume limit, the ratio of the bound state energy to that of the lowest-energy scattering state is  $\sqrt{3/4} = 0.866\ldots$.

\section{Quantum kink mass}
\label{sec:qkm}
In this section we review existing approaches to calculating the kink mass. One can calculate the leading quantum correction to the classical kink mass in the weak-coupling limit $\lambda/m^2\ll 1$ analytically by summing up the zero-point energy contributions from all the excitations (\ref{eq:goldstoneenergy}-\ref{eq:scatteringenergy}). This corresponds to one-loop level in perturbation theory, and gives~\cite{Dashen:1974cj}
\begin{equation}
\label{eq:oneloopmass}
 M_{\mathrm{1loop}} = M_{\mathrm{cl}} + m\left[ \frac{1}{6}\sqrt{\frac{3}{2}} - \frac{3}{\pi}\sqrt{2} + \mathcal{O}(\lambda/m^2) \right].
\end{equation}
This result has recently been generalised to finite volume in Refs.~\cite{Mussardo:2005dx,Pawellek:2008st}.

A non-perturbative alternative is to consider the response of the system to a `twist'~\cite{Groeneveld:1980tt}. The mass of a topological defect is then defined as the difference between the free energy $F_\mathrm{tw}$ in the topologically nontrivial sector containing the defect (in this case a kink), and the free energy $F_0$ in the topologically trivial sector,
\begin{equation}
M_{\mathrm{k}} = \frac{1}{T}\left( F_{\mathrm{tw}} - F_0 \right) = \frac{1}{T} \log \frac{Z_0}{Z_\mathrm{tw}},
\end{equation}
where $T$ is the length of the system in the time direction. Unfortunately this is impossible to measure in Monte Carlo simulations because one only samples the partition function. We therefore resort to calculating derivatives of this mass with respect to a parameter $g$ in the Lagrangian:
\begin{equation}
\label{eq:derivative}
\frac{\partial M_\mathrm{k}}{\partial g} = \frac{1}{T}\left[ \left<\frac{\partial S}{\partial g}\right>_\mathrm{tw} - \left<\frac{\partial S}{\partial g}\right>_0\right] .
\end{equation}
One then integrates this expression from a location where the mass of the defect is known exactly, such as the phase transition where the kink mass vanishes. In this paper we will take parameter $g=m^2$, and then (\ref{eq:derivative}) becomes

\begin{equation}
\label{eq:twistresp}
\frac{\partial M_\mathrm{k}}{\partial m^2} = \frac{L}{2}\left[ \langle\phi^2\rangle_\mathrm{tw} - \langle\phi^2\rangle_0\right]
\end{equation}

Unfortunately, linear error propagation is not very reliable for estimating the error in this integral; indeed, if we add more points this will tend to \emph{increase} the error without any justifiable cause. One must also consider the error resulting from the quadrature. In practice, errors in the kink mass are under better control if one uses finite differences~\cite{Rajantie:2005hi}; error propagation can then use the standard linearised results. We have then
\begin{equation}
\label{eq:fdresp}
 M(m_2^2) - M(m_1^2) = -\frac{1}{T} \ln \frac{\left< e^{-\frac{1}{2}(m_2^2-m_1^2)\sum_x \phi^2}\right>_{m_1^2,\mathrm{tw}}}{\left< e^{-\frac{1}{2}(m_2^2-m_1^2)\sum_x \phi^2}\right>_{m_1^2,\mathrm{0}}},
\end{equation}
a formula which resembles the `resampling' technique of Ferrenberg and Swendsen~\cite{Ferrenberg:1988yz}. Indeed, one way of checking that the measurement spacing is appropriate is to check that the measurements for the change of $M$ from $m_1^2$ to $m_2^2$ are the same when taken from above or below.

An equivalent criterion, applicable to (\ref{eq:twistresp}), is to actually use the `resampling' method to ensure that neighbouring values of $\langle\phi^2\rangle$ agree under the shift of $m^2$. Unfortunately, this does not remove the problem with propagation of errors.

One of the main drawbacks of this approach is the requirement to numerically integrate (or take finite differences) from, for example, the phase transition. Close to the phase transition, the errors in the expectation values in (\ref{eq:twistresp}) will be quite large due to critical slowing down. Worse, the overall error in the kink mass will increase with the length of the integration path. A local calculation of the mass of a topological defect will give better control of the errors and provide a check on the results obtained by the traditional approach.

\section{Correlation functions}
\label{sec:correlationfunctions}

In this section we consider correlation functions in the topologically nontrivial sector. Specifically, we will work with two- and four-point functions of the scalar field. We use the two-point functions to study the kink mass, and the four-point functions to study the kink spectrum. These approaches are motivated by the spectral expansion of the field in the presence of a kink.

\subsection{Spectral expansion}
\label{sec:spectralexpansion}

The correlation function of general operators ${\cal O}_i(t)$ localised in time has a spectral expansion
\begin{equation}
\label{eq:spectral}
C_{ij}(t_2-t_1) = \langle \mathcal{O}_i(t_1) \mathcal{O}_j(t_2) \rangle = \sum_{\alpha=0}^\infty \langle 0 | \mathcal{O}_i | \alpha \rangle\langle \alpha | \mathcal{O}_j | 0 \rangle e^{-(t_2-t_1)E_{\alpha}}.
\end{equation}
where is $|0\rangle$ is the ground state of the relevant topological sector and $|\alpha\rangle$ are a complete set of states with energies $E_\alpha$. By a suitable choice of operators, we can determine individual terms in this expansion. By calculating the corresponding correlator in the presence of a kink, we can therefore determine its complete excitation spectrum, at least in principle.

In practice, the energies $E_\alpha$ can be determined either by a straightforward fit to (\ref{eq:spectral}) or by using the L\"{u}scher-Wolff method~\cite{Caselle:1999tm,Gockeler:1994rx,Luscher:1990ck}. In the latter approach, we consider the generalised eigenvalue problem
\begin{equation}
\label{eq:gep}
C_{ij}(t)\rho^{(n)} = \lambda_n (t,t_0) C_{ij}(t_0) \rho^{(n)}
\end{equation}
where the eigenvalues $\lambda_n$ have the long-distance behaviour $\lambda_n(t) = e^{-t E_n}$ as $t\to \infty$. The energies are then obtained with
\begin{equation}
\label{eq:lwdecay}
E_n = \log \left(\frac{\lambda_n (t,t_0)}{\lambda_n (t+1,t_0)} \right).
\end{equation}

We show in Sections \ref{sec:twopointfunction} and \ref{sec:fourpointfunction} 
that by considering operators ${\cal O}_i$ constructed from one or two field operators $\phi$, we can measure not only the excitation spectrum in the rest mass of the kink, which corresponds to the linearised results in (\ref{eq:goldstoneenergy}-\ref{eq:scatteringenergy}), but also approximate wave functions of these states and the mass of the kink itself.

\subsection{Two-point function}

\label{sec:twopointfunction}

Let us first consider operators ${\cal O}_k=\phi(k)=\int dx \: e^{ikx}\phi(x)$ for different momenta $k$, which are quantised because of the antiperiodic boundary conditions,
\begin{equation}
k=\frac{(2n+1)\pi}{L}.
\end{equation}
Momentum conservation requires the momenta of the two operators to be equal, so for each momentum $k$ we have the correlation function
\begin{equation}
\label{eq:correlatorkinkmass}
C(t_2 - t_1; k) = \int_{x_1}\int_{x_2} d x_1 \: d x_2 \: e^{ik(x_1-x_2)} \langle \phi(x_1, t_1) \phi(x_2, t_2) \rangle; \qquad k = \frac{(2n+1)\pi}{L}
\end{equation}
as calculated in our Monte Carlo simulation for the kink sector of the broken phase. This has a spectral expansion
\begin{equation}
 C(t_2 - t_1; k) = \sum_{\alpha=1}^{\infty} \langle 0 | \phi(t_1;k) | \alpha \rangle\langle \alpha | \phi(t_2;-k) | 0 \rangle e^{-(t_2-t_1)E_{\alpha}}.
\end{equation}
Because of momentum conservation, all states $|\alpha\rangle$ in this expansion must have overall momentum $k$. The lightest such state corresponds to the boosted kink. Other states in the expansion correspond to excited states of the kink and two-particle states consisting of a kink and a scalar particle. In the limit $k\ll m$, there should be a significant gap between the energies; the correlator (\ref{eq:correlatorkinkmass}) is therefore dominated by the boosted kink at long time separation $t_2-t_1$. By analogy with the eigenspectrum given in Section \ref{sec:boundstates}, we expect

\begin{equation}
\begin{split}
\label{eq:rewrittenspectral}
C(t_2 - t_1;k)  = & \langle 0 | \phi(t_1; k) | \hbox{kink}_k \rangle \langle \hbox{kink}_k | \phi(t_2; -k) | 0 \rangle e^{-(t_2-t_1)E^{\rm kink}_{k}} + \\
& + \langle 0 | \phi(t_1; k) | \hbox{bs} \rangle \langle \hbox{bs} | \phi(t_2; -k) | 0 \rangle e^{-(t_2-t_1)E^{\rm bs}_{k}} + \\
 & + \sum_{\alpha=1}^{\infty} \langle 0 | \phi(t_1; k) | \alpha \rangle\langle \alpha | \phi(t_2; -k) | 0 \rangle e^{-(t_2-t_1)E^{\rm bulk}_{\alpha,k}}
\end{split}
\end{equation}

\FIGURE{\centering
\includegraphics[scale=0.3]{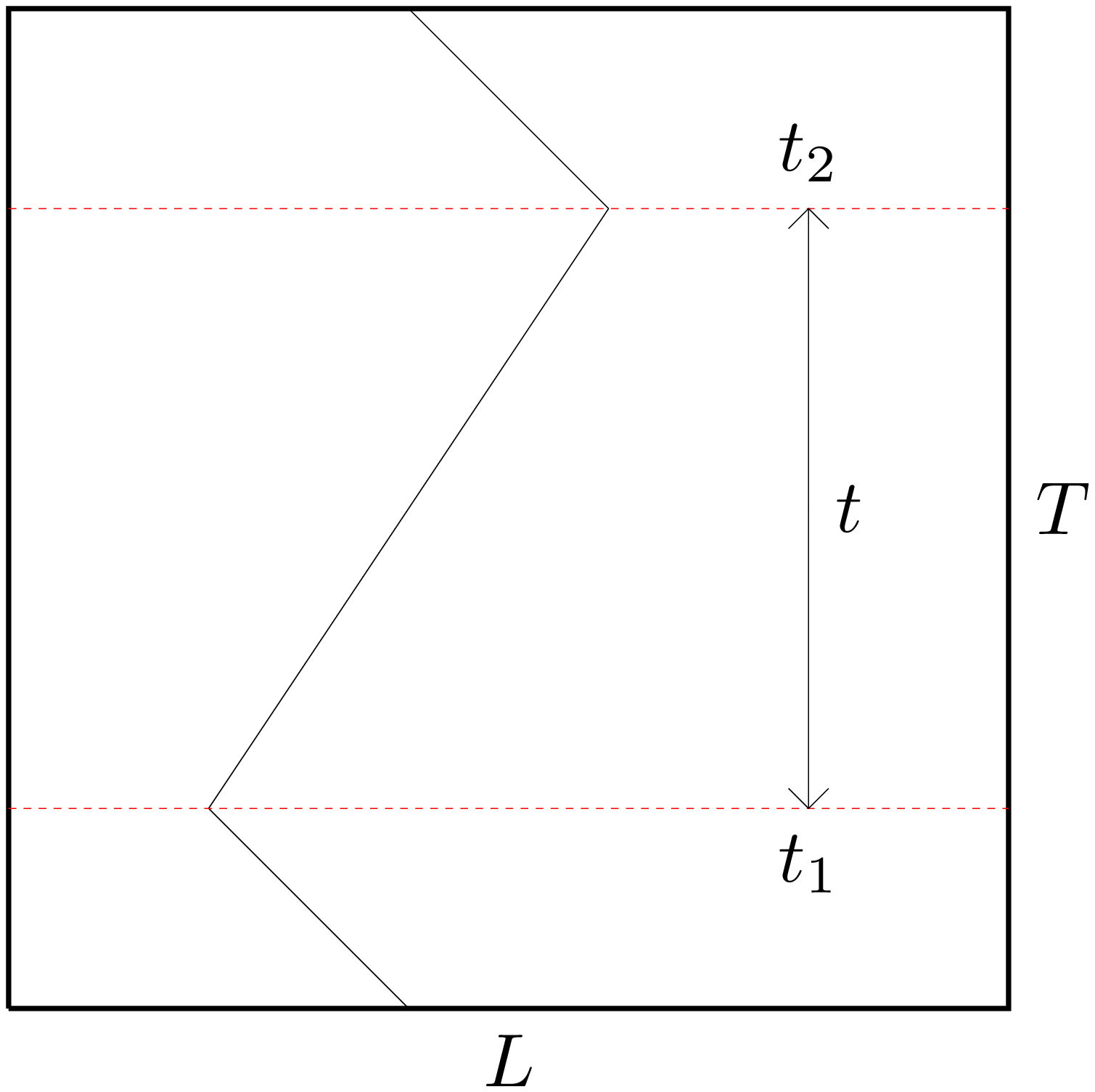}
\caption{\label{fig:cartoon} Schematic picture of `recoil' behaviour at a classical level. The kink is forced to match up due to periodic boundary conditions in the time direction (length $T$). It receives a kick from the scalar particle at $t_1$ which, to conserve momentum, it must give up at $t_2= t_1 + t$.}}
\noindent
where $E_k^{\rm kink}$ is the energy of a boosted kink with momentum $k$; $E_k^{\rm bs}$ and $E_{\alpha,k}^{\rm bulk}$ are the energies for the bound state and a scattering state (a particle in the bulk) respectively. This spectrum has only one massive bound state, but we do not rely on this assumption in the calculations which follow.

The physical interpretation of this is that the dominant contribution to the correlator comes from the $\phi$ particle created at $t_1$ absorbed by the kink before re-emerging to be annihilated at $t_2$. Not only must momentum be conserved by this process, but we also impose periodicity in the time direction: $\phi(x,t)=\phi(x,T+t)$. As long as $k\ll m$ and $(t_2-t_1)\gg 1/m$, the kink width is small relative to other relevant length scales, we can treat the kink as a point particle. Close to the phase transition, $m$ approaches zero and the kink width diverges, and lower momenta $k$ and longer time separations have to be used.

Consider the action of a point particle of mass $M$ in free space whose momentum increases by $k$ at $t_1$ and decreases by $k$ at $t_2=t_1+t$. Periodic boundary conditions require that the initial and final $x$-components of the velocity must match. Then,
\begin{equation}
S = M\left[t \sqrt{1+v_1^2} + (T-t)\sqrt{1+v_0^2} \right]
\end{equation}
noting that $v_0$ and $v_1$ are related by $t \leftrightarrow (T-t)$. We assume that the kink moves non-relativistically, so $v_0,\: v_1 \ll 1$ and obtain
\begin{equation}
 S \approx  M\left[T + \frac{1}{2}v_1^2 t + \frac{1}{2} v_0^2 (T-t)\right]
\end{equation}
and, by applying the stationary phase approximation we expect to see a dominant contribution to the correlation function, as measured on the lattice, that has the form
\begin{equation}
\label{eq:goldstone}
C(t;k)_\mathrm{Gs} \sim \exp\left[-\frac{1}{2}\frac{k^2}{M}\frac{ t (T - t)}{T}\right].
\end{equation}
In the infinite-time limit, $T\rightarrow\infty$, this reduces to the non-relativistic kinetic energy of a particle with momentum $k$~\cite{Goldstone:1974gf}. In practice, the finite $T$ effects are important, and therefore we keep the equation in this form. By fitting the asymptotic long-time behaviour of the correlator (\ref{eq:correlatorkinkmass}) to Equation (\ref{eq:goldstone}), we can determine the kink mass $M$ because the momentum $k$ is fixed. This gives us a direct way of measuring the kink mass.

\subsection{Four-point functions and excitation spectrum}
\label{sec:fourpointfunction}

To find the excitation spectrum in the rest frame of the kink, we need to use operators with zero overall momentum. Therefore we consider two-particle operators which correspond to creating two particles on a timeslice, separated by distance\footnote{One could also work with Fourier-transformed fields with appropriate antiperiodic momenta.} $\Delta x$.

\begin{equation}
\label{eq:twoparticle}
\mathcal{O}_{\Delta x} (t) = \sum_{x} \phi(x,t)\phi(x+\Delta x,t)
\end{equation}
We can use the L\"{u}scher-Wolff method (\ref{eq:lwdecay}) to determine the energy spectrum $\{E_\alpha\}$, which now includes only states in the rest frame of the kink and should therefore correspond directly to the negative parity states in (\ref{eq:goldstoneenergy}-\ref{eq:scatteringenergy}). Whether a given state is a localised bound state of the kink or a free particle state can be determined by investigating its dependence on the volume $L$. In the infinite-volume limit, free particle states should behave as
\begin{equation}
\label{eq:dispersion}
E^2\sim m^2+k^2=m^2+\frac{(2n+1)^2\pi^2}{L^2}+O(L^{-4}).
\end{equation}
In contrast, a localised bound state would have an exponentially small finite-size effect. When the volume $L$ is smaller than the inverse mass $1/m$ we would expect the kink to be `squeezed' and the bound state to have lower energy.

The other way of seeing the localisation of the first state is to try to reconstruct the wavefunctions from the generalised eigenvectors of (\ref{eq:gep}). We can then compare our results with the eigenfunctions of the fluctuation field in the presence of a continuum kink (\ref{eq:bschr}), or its discrete equivalent discussed later.

Using (\ref{eq:spectral}) to rewrite the generalised eigenvalue problem (\ref{eq:gep})
\begin{equation}
\sum_l \left< 0 \right| \mathcal{O}_{\Delta x} \left| l \right> e^{-E_l t} \sum_{\Delta y} \left< l \right| \mathcal{O}_{\Delta y} \left| 0 \right> \rho_{\Delta y}^{(n)} = \lambda_n \sum_m \left< 0 \right| \mathcal{O}_{\Delta x} \left| m \right> \sum_{\Delta z} \left< m \right| \mathcal{O}_{\Delta z} \left| 0 \right> \rho_{\Delta z}^{(n)},
\end{equation}
we expect $\lambda_n(t) = e^{-t E_n}$ at long distance and so
\begin{equation}
  \sum_{\Delta y} \left< m \right| \mathcal{O}_{\Delta y} \left| 0 \right> \rho_{\Delta y}^{(n)} = \delta_{mn}
\end{equation}
is a reasonable ansatz. Using our choice of two-particle operators this becomes
\begin{equation}
  \sum_y \sum_{\Delta y} \left< m \right| \phi(y) \phi(y+\Delta y) \left| 0 \right> \rho_{\Delta y}^{(n)} = \delta_{mn}.
\end{equation}
We now split the field $\phi$ into the kink background $\phi_k$ and fluctuations $\hat{\phi}$. We are in the broken phase; we assume in our expansion that the one-particle states are well-separated from the two-particle states. Then for the one-particle states
\begin{equation}
   \sum_y \sum_{\Delta y} \left< m \right| \left[\phi_k(y+\Delta y) + \phi_k(y-\Delta y)\right] \hat{\phi}(y) \left| 0 \right> \rho_{\Delta y}^{(n)}= \delta_{mn}.
\end{equation}
We further assume that the field $\hat{\phi}$ can be decomposed into orthogonal modes -- the wavefunctions for each energy eigenstate. Noting that this means
\begin{equation}
 \sum_x \hat{\phi}_m (x) \hat{\phi}_n (x) = \delta_{mn}
\end{equation}
we identify
\begin{equation}
 \hat{\phi}_n(x) = \sum_{\Delta x} \rho^{(n)}_{\Delta x}\left[\phi_k (x+\Delta x) + \phi_k (x-\Delta x)\right].
\end{equation}

This shows that, for one-particle states, the convolution of the generalised eigenvectors $\rho^{(n)}$ of the correlation matrices $C(t)$, $C(t_0)$ with the kink background gives the wavefunction. A starting assumption for the form of the kink background in the weak-coupling limit is that it takes the standard shape (\ref{eq:kinksolution}) but with the renormalised mass taking the place of $m$. This can be obtained from simulations with periodic boundary conditions.

\section{Numerical calculations and simulations}
\label{sec:numerics}

For calculations on a lattice of size $L\times T$, we make the standard choice of discretisation:
\begin{equation}
\label{eq:latticeaction}
S = \sum_{\mathbf{x}} \left[ -\sum_{\mu=1}^2 \phi(\mathbf{x})\phi(\mathbf{x}+a \mathbf{\hat{\mu}}) +  a^2\left(2 + \frac{m^2}{2}\right)\phi(\mathbf{x})^2 + a^2\frac{\lambda}{4!}\phi(\mathbf{x})^4 \right]
\end{equation}
where the summation over $\mathbf{x}$ runs over all sites, and $\mathbf{\hat{\mu}}$ is a unit vector in either direction on the lattice. For the remainder of this paper, we have set $a=1$, and instead vary $\lambda$ and $m^2$.

Periodic boundary conditions are employed in the time direction at all times. The topologically nontrivial sector, where the kink is present, is simulated by imposing antiperiodic boundary conditions on the lattice in the spatial direction.

Finite lattice spacing changes the dispersion relation by introducing a momentum cutoff. For a free particle, the dispersion relation is
\begin{equation}
E(k)=\sqrt{\frac{4}{a^2}\sin^2\frac{ak}{2}+m^2}.
\end{equation}
This modifies the $O(L^{-4})$ contribution in Equation (\ref{eq:dispersion}).

\subsection{Linearised calculations}
\label{sec:lincalc}

The linearised energy spectrum in Equations (\ref{eq:goldstoneenergy}-\ref{eq:scatteringenergy}) is calculated in continuum, and to be able to compare it with our Monte Carlo results, which are obtained on a discrete lattice, we need to understand what effect the discretisation itself has. We can see this by considering the Schrodinger problem for a kink in discrete space.

A straightforward way of understanding the finite-size and discretisation effects is to solve the lattice Schr\"{o}dinger equation for a kink background. Splitting $\phi$ into kink background $\phi_k$ and fluctuations $\hat{\phi}$ as before, we find the classical discrete kink by the gradient descent method,

\begin{multline}
\label{eq:gradientflow}
\phi_k^{(n+1)}(x) = \phi_k^{(n)}(x) - \Delta t \Bigg[2\phi_k^{(n)}(x) - \phi_k^{(n)}(x+a) -\\
 - \phi_k^{(n)}(x-a) - m^2 \phi_k^{(n)}(x) + \frac{\lambda}{3!} \left(\phi_k^{(n)}(x)\right)^3 \Bigg]
\end{multline}
where antiperiodic boundary conditions are used and $\Delta t$ is chosen to give good convergence. The eigenvectors and eigenvalues of $\hat{\phi}$ are then obtained by solving
\begin{equation}
\label{eq:latticeschr}
\left[\mathsf{D^2} - \frac{m^2}{2}\mathbbm{1} + \frac{\lambda}{4}\mathsf{B}\right]\phi_n(x) = E^2 \phi_n(x)
\end{equation}
where $\mathsf{D^2}$ is the second-order lattice derivative with antiperiodic boundary conditions, and $\mathsf{B}$ is the diagonal matrix with entries $\mathsf{B}_{xx}=\phi_k (x)^2$.

\FIGURE{
\centering
\includegraphics[scale=0.4]{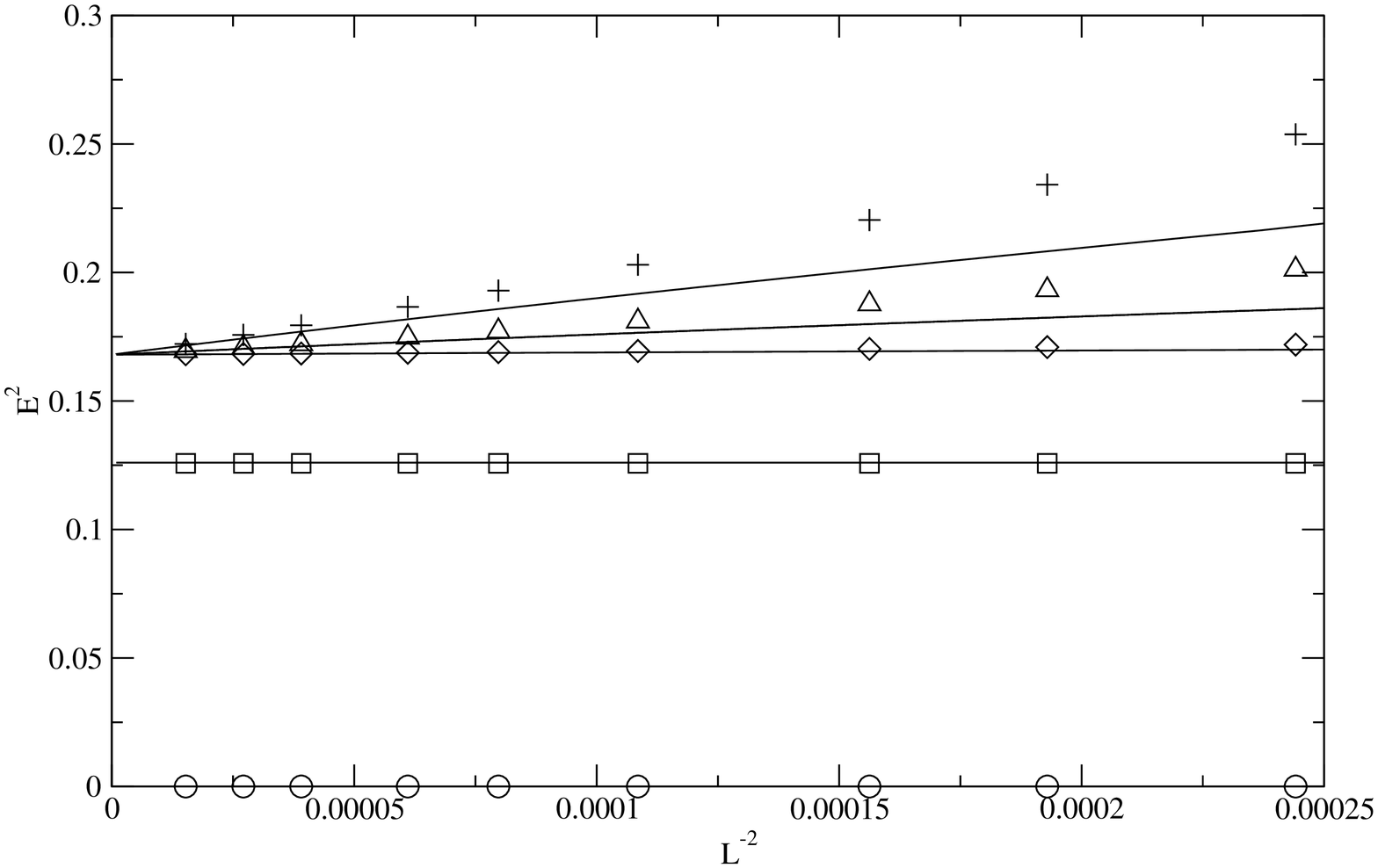}
\caption{\label{fig:splot} Plot of the eigenvalues $E^2$ of the discretised Schr\"{o}dinger equation (\ref{eq:latticeschr}) compared with the corresponding results for the continuum with a finite box, for $\lambda=1/16$, $m^2=0.084$. Plotted are the Goldstone mode (circles), the bound state (squares) and the first three scattering states (diamonds, triangles, crosses). There is a significant finite-size behaviour at this choice of $m^2$, which will be worsened by radiative corrections.}
}

The eigenvalues are compared with the continuum result (\ref{eq:allowedant}) in Figure \ref{fig:splot} for $m^2=0.084$, $\lambda=1/16$. These cannot be directly compared with the Monte Carlo results as the mass will pick up radiative corrections, but serve to give a qualitative description of the discretisation effects.

\subsection{Monte Carlo simulations}

Our Monte Carlo simulation uses a standard Metropolis algorithm, with acceptance rates at around 70\%. A hybrid Monte Carlo algorithm was tried but not used for the results presented here, as it did not deliver a significant overall improvement in performance. Generally, the space between measurements was kept longer than the autocorrelation time for the observable of interest. For the L\"{u}scher-Wolff eigenvectors, it is impossible to determine the autocorrelation time but it was assumed to be far longer than that for the two-point function results.

\subsubsection{Results for the kink mass}

As a benchmark, we first computed the kink mass using the conventional twist method (see Section \ref{sec:qkm}) for many different values of $m^2$ along the line $\lambda=1/16$ in parameter space. The spatial average of $\phi^2$ was measured for antiperiodic and periodic spatial boundary conditions, so that the mass could be calculated from (\ref{eq:fdresp}). At least $2\times 10^5$ measurements were carried out for each value of $m^2$, separated by 50 update sweeps. The integrated autocorrelation time for $\phi^2$~\cite{Montvay1997},
\begin{equation}
\tau_{\mathrm{int},\phi^2} = \frac{1}{2}\sum_\tau \frac{\langle \phi^2_n \phi^2_{n+\tau} \rangle - \langle \phi^2_n \rangle \langle \phi^2_{n+\tau} \rangle}{\langle \phi^4 \rangle - \langle \phi^2 \rangle^2},
\end{equation}
was then estimated and used to find the number of independent measurements available for both the periodic and antiperiodic cases. The measurements were then binned appropriately. These $\phi^2$ measurements were then used to calculate
\begin{eqnarray}
 f_1 & = & -\frac{1}{T} \ln \left< e^{-\frac{1}{2}(m_2^2 - m_1^2)\sum_x \phi^2} \right>_{m_1^2}, \\
 f_2 & = & -\frac{1}{T} \ln \left< e^{-\frac{1}{2}(m_2^2 - m_1^2)\sum_x \phi^2} \right>_{m_2^2}
\end{eqnarray}
in both the kink and trivial sectors. By analogy with the resampling technique of Ferrenberg and Swendsen, the measurement spacing has been made small enough that $f_1$ and $f_2$ agree, within errors, in each sector. The final errors in the kink mass derivative were estimated by applying the bootstrap method. The kink mass is then obtained by using these data to calculate
\begin{equation}
M(m_2^2) - M(m_1^2) = \frac{1}{2}(f_{1,\mathrm{tw}} + f_{2,\mathrm{tw}} - f_{1,\mathrm{0}} - f_{2,\mathrm{0}})
\end{equation}
where `${\rm tw}$' demotes the results of simulations with antiperiodic (twisted) boundary conditions and `0' those with periodic boundary conditions. The associated error is then
\begin{multline}
\Delta\left[ M(m_2^2) - M(m_1^2) \right]= \frac{1}{4}\left[ \Delta f_{1,\mathrm{tw}}^2 + \Delta f_{2,\mathrm{tw}}^2 + \Delta f_{1,0}^2 + \Delta f_{2,0}^2\right.  \\
\left.+ (f_{1,\mathrm{tw}} - f_{2,\mathrm{tw}})^2 + (f_{1,0} - f_{2,0})^2\right].
\end{multline}
These expressions can then be used to find the kink mass
\begin{equation}
 M(m_N^2) = \sum_{n=0}^{N-1} \left[ M(m_{n+1}^2) - M(m_n^2)\right].
\end{equation}
Standard error propagation techniques can then be used to calculate the error in $M(m^2)$.

Our new method, outlined in Section \ref{sec:twopointfunction}, can be used to calculate the mass at \emph{any} point without depending on measurements at any other. We calculate the correlator (\ref{eq:correlatorkinkmass}) on the lattice,
\begin{equation}
\label{eq:correxplain}
C ( t_1 - t_2;k ) =  \left< \sum_{x_1,x_2}e^{ik(x_1-x_2)} \phi(x_1,t_1)\phi(x_2,t_2) \right>;\qquad k=\frac{(2n+1)\pi}{L}
\end{equation}
and fit it to $A_1$, $A_2$, $M$ and $E$ in the function
\begin{equation}
C(t;k)=A_1\exp\left(-\frac{k^2}{2M}\frac{t(T-t)}{T}\right)
+A_2\left(e^{-Et}+e^{-E(T-t)}\right),
\end{equation}
where the first term corresponds to the contribution (\ref{eq:goldstone}) from a moving kink and the second term to a free scalar particle. A fit to either a free scalar particle or (\ref{eq:goldstone}) alone does not converge. The bound state of the kink cannot be determined by this fit, as it will be suppressed by a factor of $1/L$.

\FIGURE{
\centering
\includegraphics[scale=0.2]{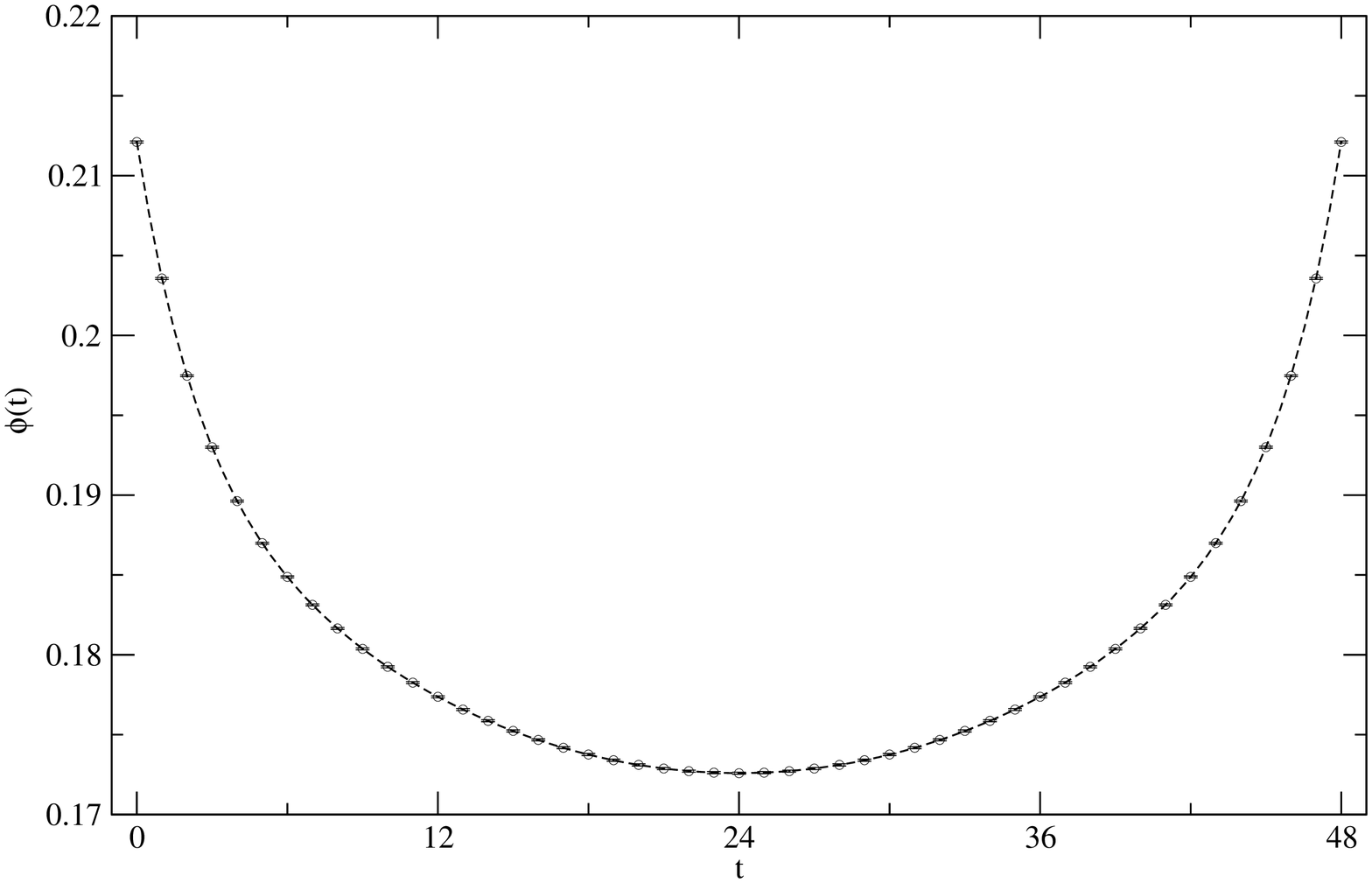}
\caption{\label{fig:examplecorrelator} Example Monte Carlo correlator data and fit for $T=48$, $m^2=0.1$, $\lambda=1/16$ with $k=3\pi/48$. Three points at each end are excluded from the fit.}
}

To exclude short-range behaviour, the fit is repeated excluding more and more short-distance measurements until a `plateau' for the fit parameters is reached. Our error estimates for the fit are obtained by performing an elaborate bootstrap: we resample our measurements of the correlator $C(t_1 - t_2;k)$, then repeat the fit to obtain a series of estimates~\cite{Efron1982}.

Our results, for varying $m^2$, are shown in Figure \ref{fig:kinkmass}. Our results agree with~\cite{Ardekani:1998tz}, in that the nonperturbative Monte Carlo result lies below the semiclassical mass result\footnote{There is an error in the equivalent calculation in~\cite{Ciria:1993yx} which was previously noted in~\cite{Ardekani:1998tz}.}. The two measurement methods are in agreement, from very close to the phase transition to beyond $M\sim a = 1$ where the treatment of the kink as a semiclassical particle might be expected to break down.  The same number of measurements were made for each parameter choice in both methods. This means that the overall number of measurements needed for any given point is greatly reduced by studying the two-point function, as the errors are comparable in the two cases.

\FIGURE{
\includegraphics[scale=0.4]{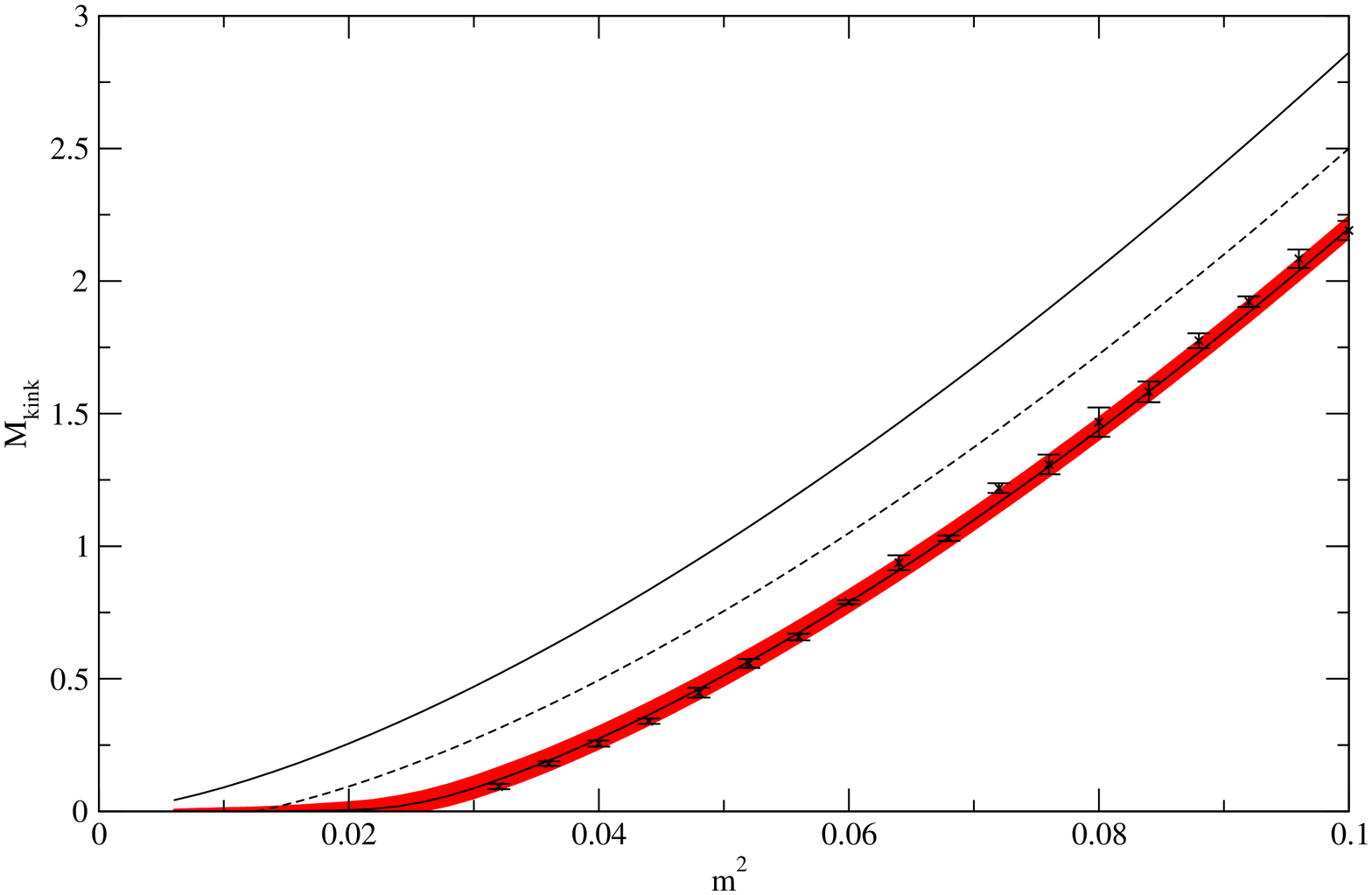}
\caption{\label{fig:kinkmass} Comparison of the kink mass measurements from a twist (Section \ref{sec:qkm}; shaded region with line at centre) and from the scattering method (Section \ref{sec:twopointfunction}; crosses). The classical (\ref{eq:classicalmass}) and first-order perturbative (\ref{eq:oneloopmass}) results are also shown as solid and dotted lines respectively. Both types of measurement were carried out for $L=48$; volumes $L=32$ and $L=40$ show finite-size effects that are similar to the magnitude of the bootstrapped errors. For the scattering measurements the errors increase when $M>1$, but the technique is still valid.}
}

\TABLE{
\centering

\begin{tabular}{| c || l | l | l | l |}
\hline
 \multicolumn{5}{|l|}{$k=\pi/L$} \\
\hline
$L$ & Kink weight $A_1$ & Bulk weight $A_2$ & E & M  \\
\hline
$32$ & $2.83875 \pm 0.00095 $ & $0.02293 \pm 0.00081 $ & $0.38382 \pm 0.01332$ & $2.22514 \pm 0.05441$  \\
$40$ & $3.02296 \pm 0.00152$ & $0.01885 \pm 0.00077$ & $0.39537 \pm 0.05527 $ & $2.29455 \pm 0.10858 $  \\
$48$ & $3.13151 \pm 0.00047$ & $0.01743 \pm 0.00019$ & $0.41412 \pm 0.00800$ & $2.19456 \pm 0.03601 $  \\
\hline
 \multicolumn{5}{|l|}{$k=3 \pi/L$} \\
\hline
$32$ & $0.09402 \pm 0.00041$ & $0.03040 \pm 0.00052$ & $0.44331 \pm 0.01196$ & $2.04267 \pm 0.07687$  \\
$40$ & $0.14710 \pm 0.00043$ & $0.02405 \pm 0.00028$ & $0.42003 \pm 0.00768$ & $2.13589 \pm 0.07250$ \\
$48$ & $0.19189 \pm 0.00044$ & $0.02016 \pm 0.00015$ & $0.41473 \pm 0.00602$ & $2.18788 \pm 0.04752$  \\
\hline
\end{tabular}
\caption{\label{tab:contributions} Contributions to the correlator (\ref{eq:correxplain}) from `kink scattering' and `bulk' behaviour for $m^2=0.1$ and $\lambda=1/16$ on a $L\times L$ lattice. The errors quoted here are from a bootstrapping method. At higher momenta the kink contribution becomes smaller, relative to the bulk contribution. However, as the lattice size is increased the kink contribution increases. The kink mass remains within the error shown in Figure \ref{fig:kinkmass} for all measurements at $k=\pi/L$.}
}

In Table \ref{tab:contributions} we show the relative contributions of `kink scattering' and `bulk' behaviour to (\ref{eq:correxplain}) for a value of $m^2$ deep in the broken phase, and a sample plot of the corresponding data is given in Figure \ref{fig:examplecorrelator}.

\subsubsection{Results for the particle spectrum in the presence of a kink}
\label{sec:matrix}

We measured the operators (\ref{eq:twoparticle}) for separations $\Delta x = 0, 1 \ldots, L/2 - 1$. An important requirement is that these operators are linearly independent; to ensure this we only use the first $L/2$ operators~\cite{Gockeler:1994rx}. This is a similar imposition in position space to that of L\"{u}scher and Wolff in momentum space for their original paper~\cite{Luscher:1990ck}. Our next step was to solve the generalised eigenvalue problem (\ref{eq:gep}) to give the energy spectrum.

\FIGURE[b]{
\centering
\includegraphics[scale=0.4]{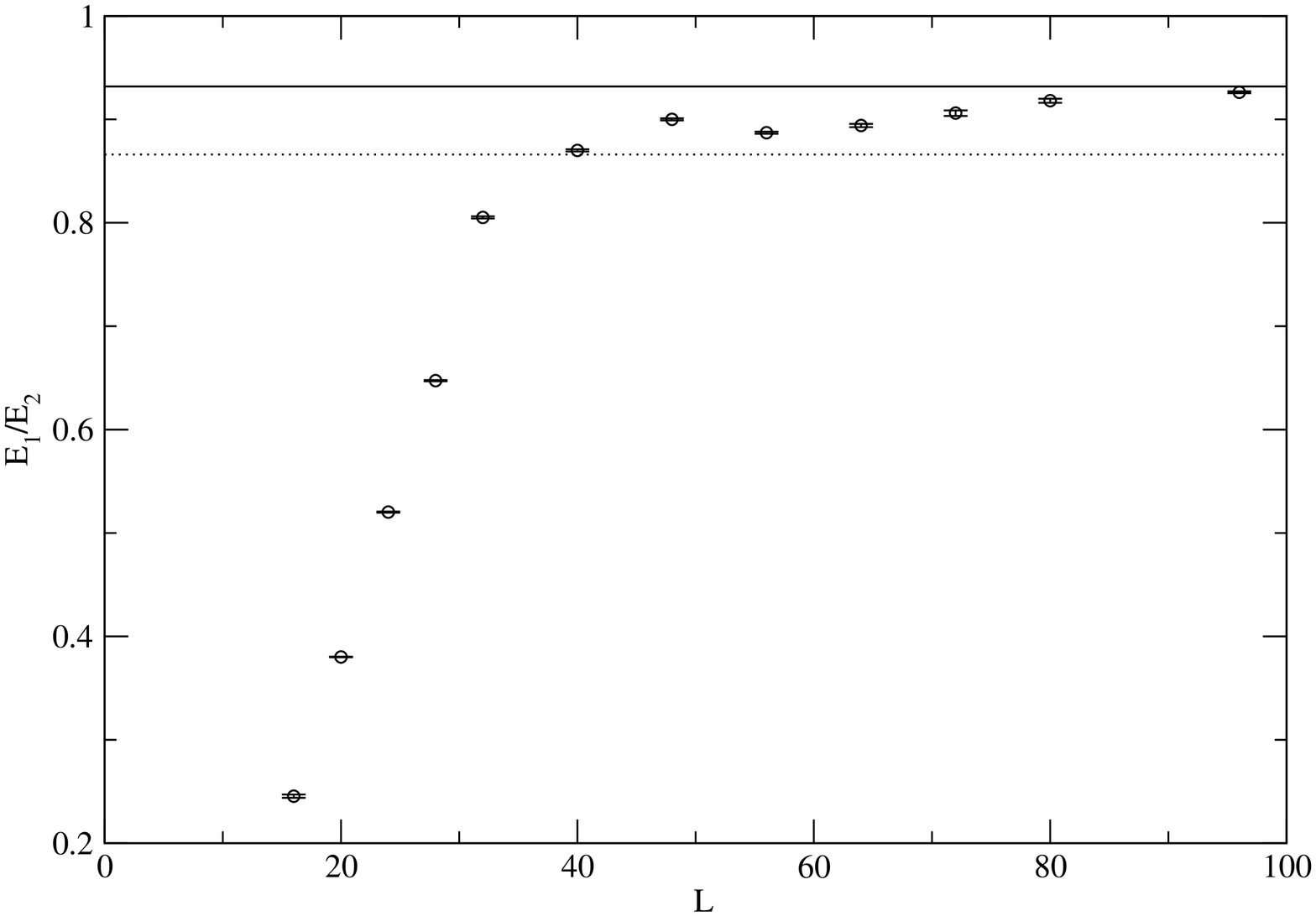}
\caption{\label{fig:ratio} Plot of the ratio of the first to the second energy eigenvalue against lattice size, for $\lambda=1/16$, $m^2=0.084$. The classical continuum result $m_{\mathrm{bs}}/m_\phi$ is shown (dotted line), as well as the estimated continuum result obtained by the intercepts of Figure \ref{fig:finitesize0084} (solid line). The results here are at $t=3$.}
}

To find the maximum time separation $t$ at which these measurements can be taken, we used the `self-adjusting exponential fit' of G\"{o}ckeler et al.~\cite{Gockeler:1994rx}, which compares sorting by largest eigenvalue at successive times $(t,t+1)$ to sorting by magnitude of the scalar product of eigenvectors. The two sorting procedures no longer agree at a time $t$ where the data are too noisy to be reliable. Another check on the extent of the noise present is the asymmetry of $C_{\Delta x,\Delta y}$. The noise in the data readily becomes apparent when fits or wavefunction reconstruction are attempted but these procedures serve as valuable checks.

In order to verify the presence of a bound, localised state in the theory, simulations were carried out on lattices of spatial size $L$ and time size $3L$ at various lattice sizes, to see if the squeezing of the kink and the bound state could be observed. In these simulations, $\lambda=1/16$ was used, with $m^2=0.084$ and $m^2=0.504$. These parameters have been chosen so that $\lambda/m^2$ is slightly less than unity while $1/m$ is large enough to see the squeezing of the kink at smaller lattice sizes. This squeezing effect is seen in the ratio of the bound state energy to that of the first scattering state, plotted in Figure \ref{fig:ratio}.

The finite-size effects and the phase shift due to the kink can be seen when we attempt to extrapolate to the continuum limit (Figures \ref{fig:finitesize0084} and \ref{fig:finitesize0504}). These should be contrasted with the Schr\"{o}dinger results of Figure \ref{fig:splot}.

\FIGURE{
\centering
\includegraphics[scale=0.4]{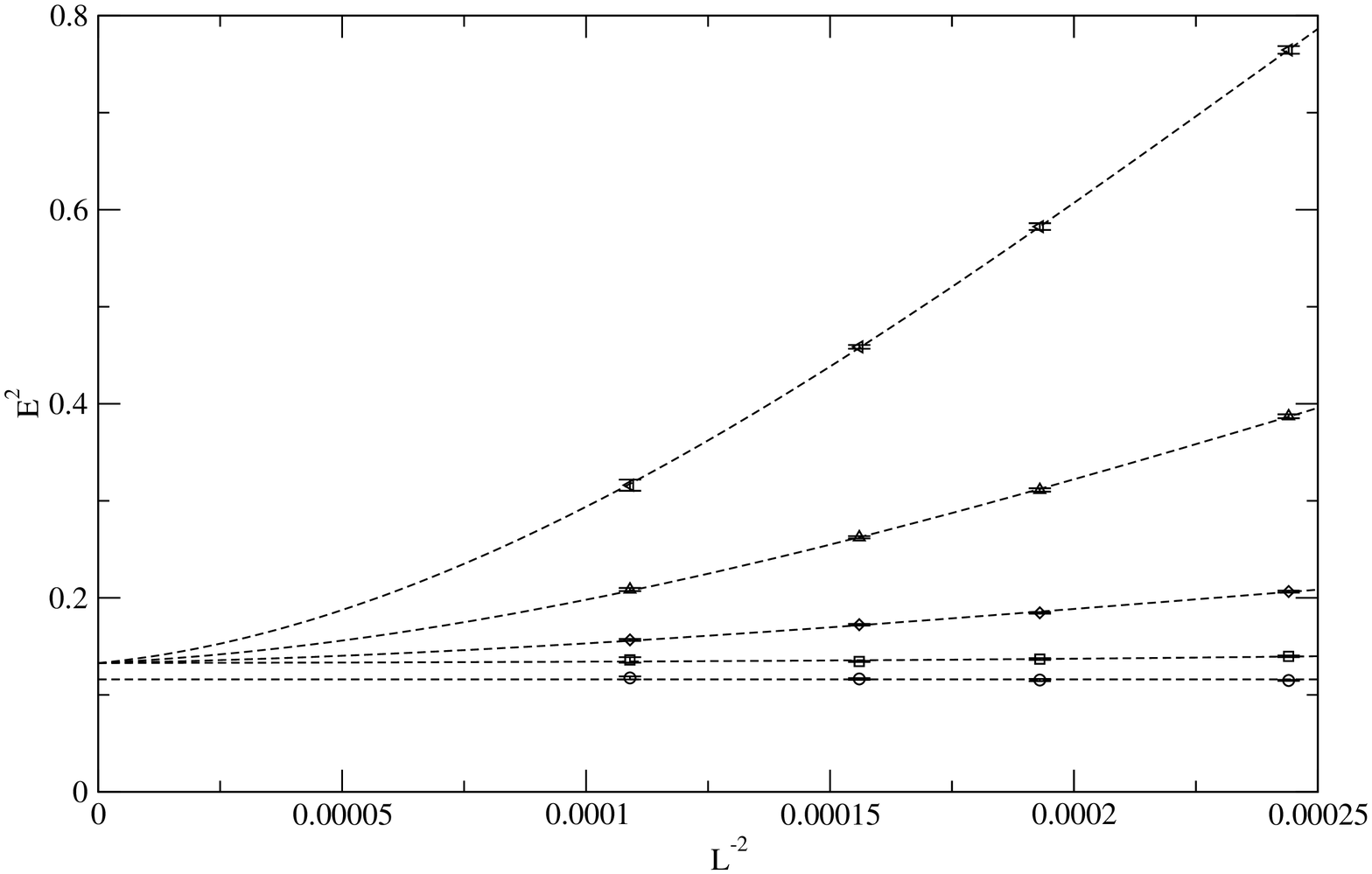}
\caption{\label{fig:finitesize0084} Plot of the first five energy levels given by the diagonalisation of (\ref{eq:spectral}) for the operators (\ref{eq:twoparticle}), with $\lambda=1/16$, $m^2=0.084$. This was done at relatively short distance ($t=2$). For the scattering states, the quality of the data here has allowed a constrained fit with two orders of discretisation effects (\ref{eq:thefit}).}
}

\FIGURE{
\centering
\includegraphics[scale=0.4]{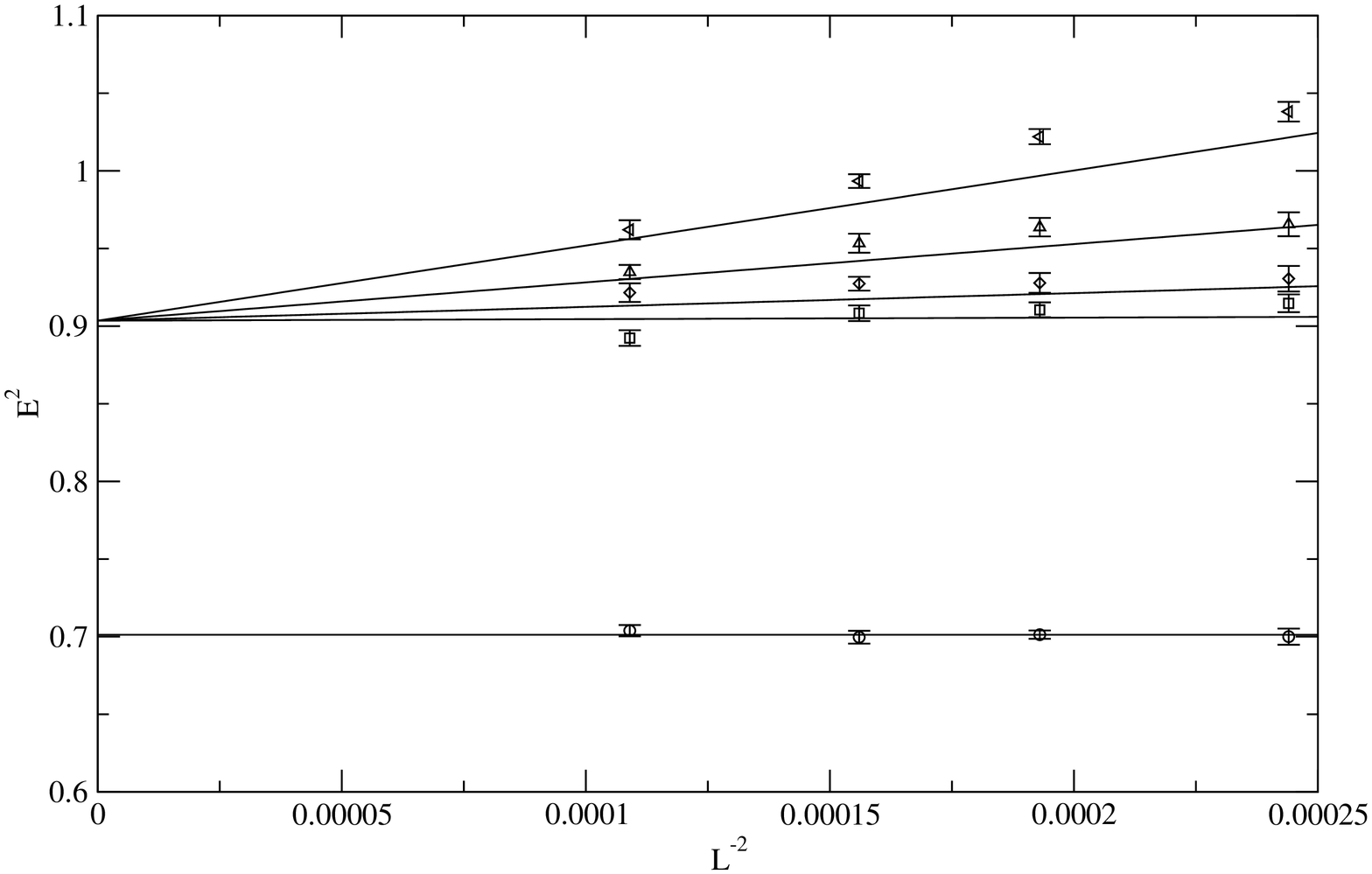}
\caption{\label{fig:finitesize0504} As Figure \ref{fig:finitesize0084} but with $\lambda=1/16$, $m^2 = 0.504$. Here the lines are not fits, but show the asymptotic free-particle behaviour $E^2 = m^2 + (2n+1)^2\pi^2/L^2$.}
}

For both plots, the lowest energy level is fit to a constant since it should not have any volume dependence. For the other excitations we can include the finite-size effects. From (\ref{eq:dispersion}), this motivates a fit to $c_1$ and $c_2$ in
\begin{equation}
\label{eq:thefit}
E^2 = m^2 + \frac{(2n+1)^2\pi^2}{L^2} + c_1 \frac{1}{L^4} + c_2 \frac{1}{L^6}
\end{equation}
for Figure \ref{fig:finitesize0084}. The value of $m^2$ is obtained from simulations with periodic boundary conditions. The data in Figure \ref{fig:finitesize0504} are, however, too noisy to permit the same treatment. Instead we show the asymptotic behaviour given by the first two terms.

\subsubsection{Eigenvectors and wavefunctions}
In addition to the finite-size behaviour, further evidence for the bound state is given by the shape of the eigenvectors. In Figures \ref{fig:l64evs084} and \ref{fig:l64evs504}, the eigenvectors of the correlation matrix are given for the first four energy levels at $L=64$ (the error bars are estimates based on a bootstrap of the eigenvectors for resampled sets of measurements).

\FIGURE[p]{
\centering
\includegraphics[scale=0.4]{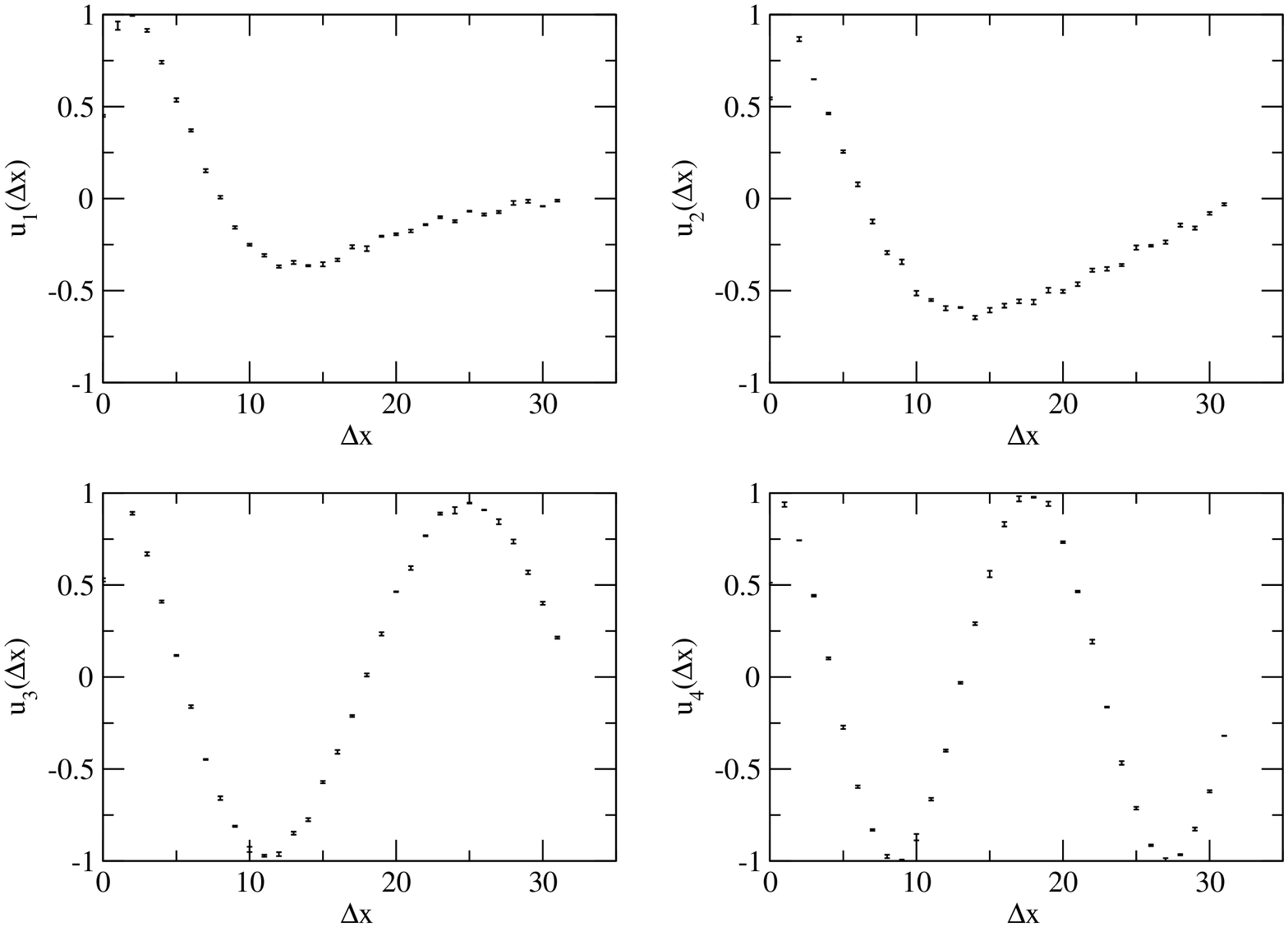}
\caption{\label{fig:l64evs084} Plot of four lowest-lying eigenvectors for $L=64$, $m^2=0.084$.}
}

\FIGURE[p]{
\centering
\includegraphics[scale=0.4]{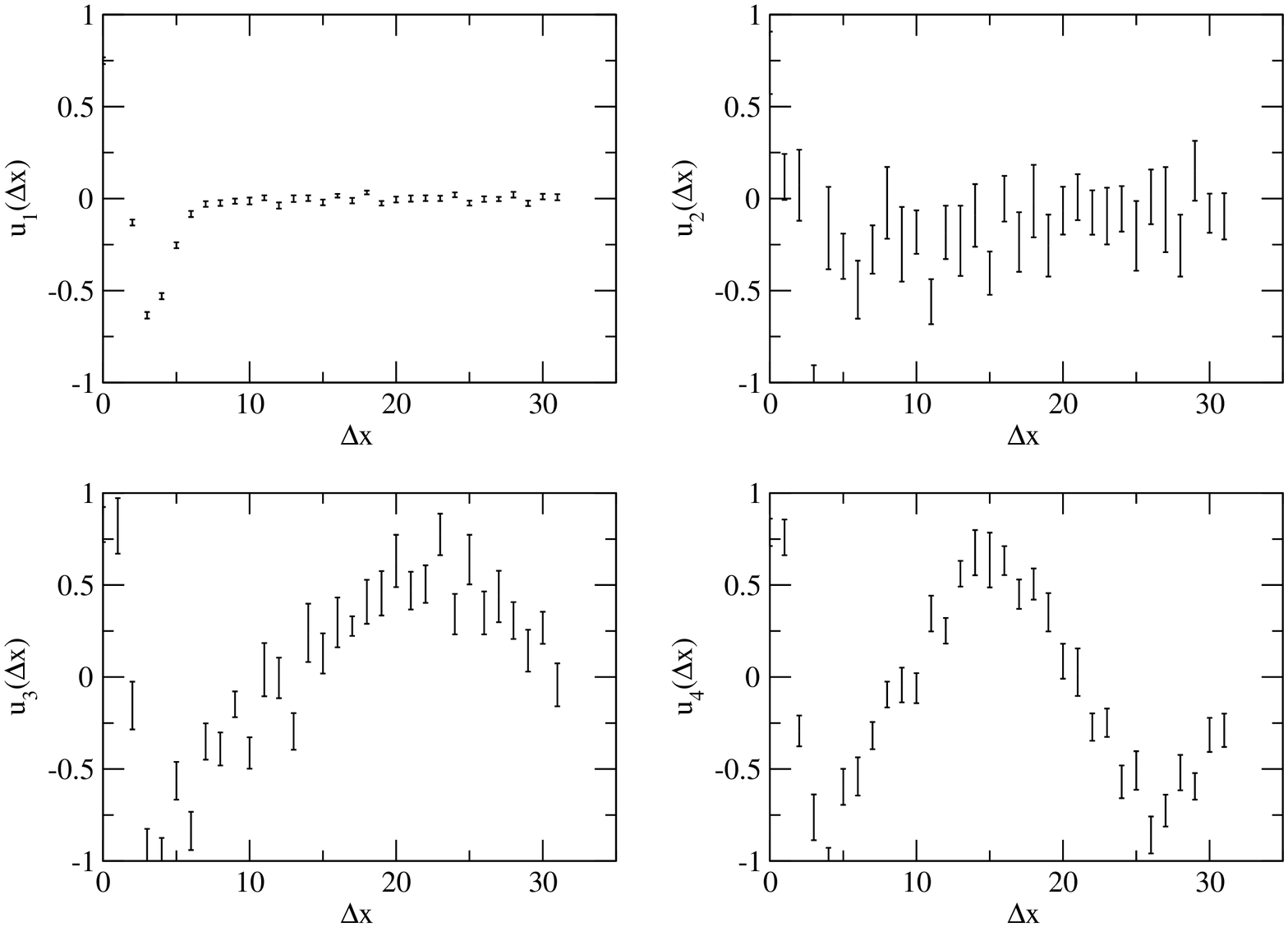}
\caption{\label{fig:l64evs504} Plot of four lowest-lying eigenvectors for $L=64$, $m^2=0.504$.}

}

Following the method given in Section \ref{sec:fourpointfunction}, these eigenvectors can be used to reconstruct the original wavefunctions. First, we estimate the renormalised mass by fitting the zero-mode energies in the periodic sector to a constant. This renormalised mass is then used in the classical gradient flow (\ref{eq:gradientflow}) method to estimate the background kink. 

Approximate errors are recalculated using the bootstrap method; interestingly, the errors are much smaller than those for the corresponding eigenvector. The results of the reconstruction are shown in Figures \ref{fig:l64evs084} and \ref{fig:l64evs504}. In Figure \ref{fig:l64evs084} the kink is slightly wider than the lattice spatial size, and hence the bound state is `squeezed'.

\FIGURE[p]{
\centering
\includegraphics[scale=0.4]{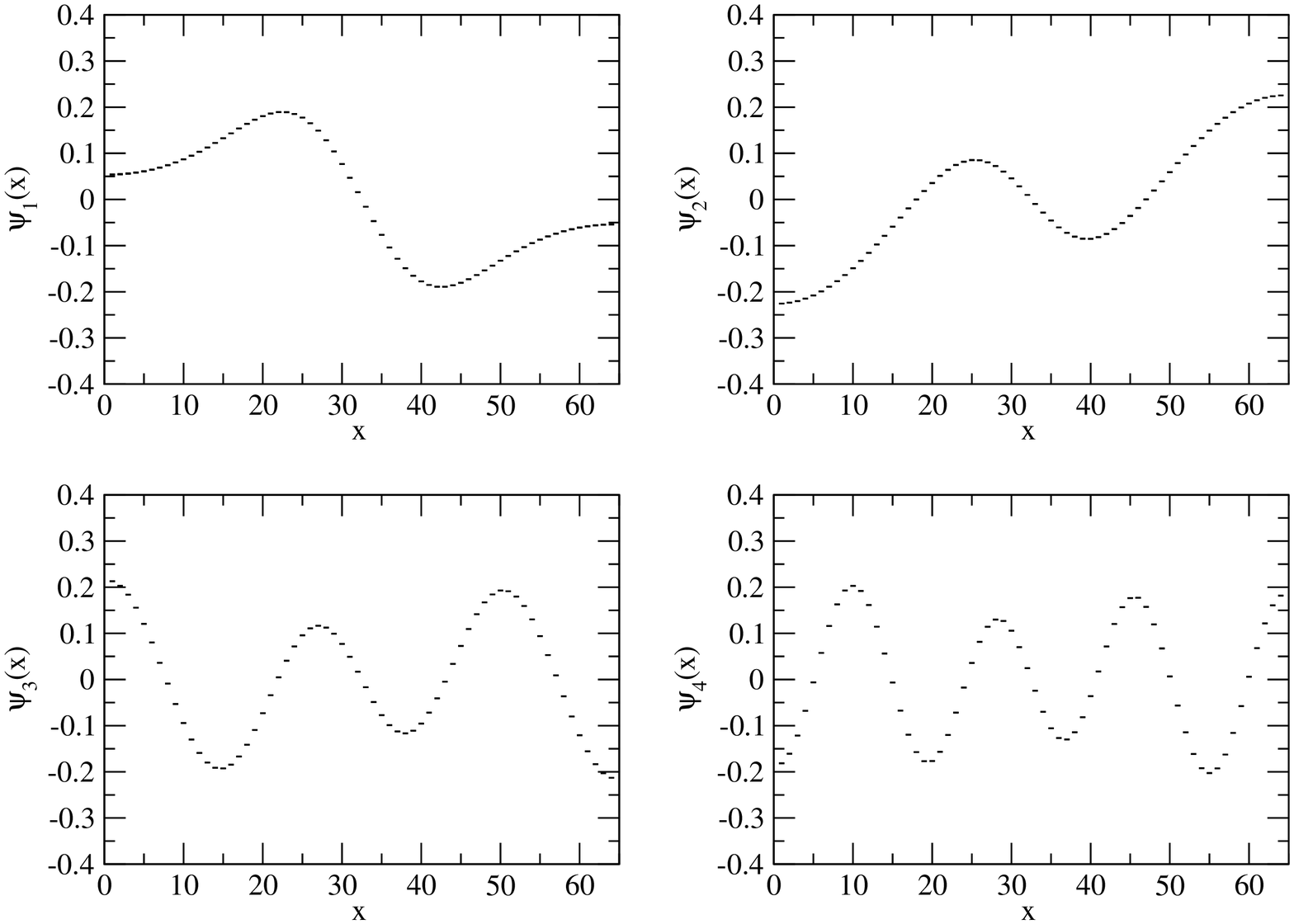}
\caption{\label{fig:l64wfs084} Plot of four lowest-lying wavefunctions for $L=64$, $m^2=0.084$.}
}

\FIGURE[p]{
\centering
\includegraphics[scale=0.4]{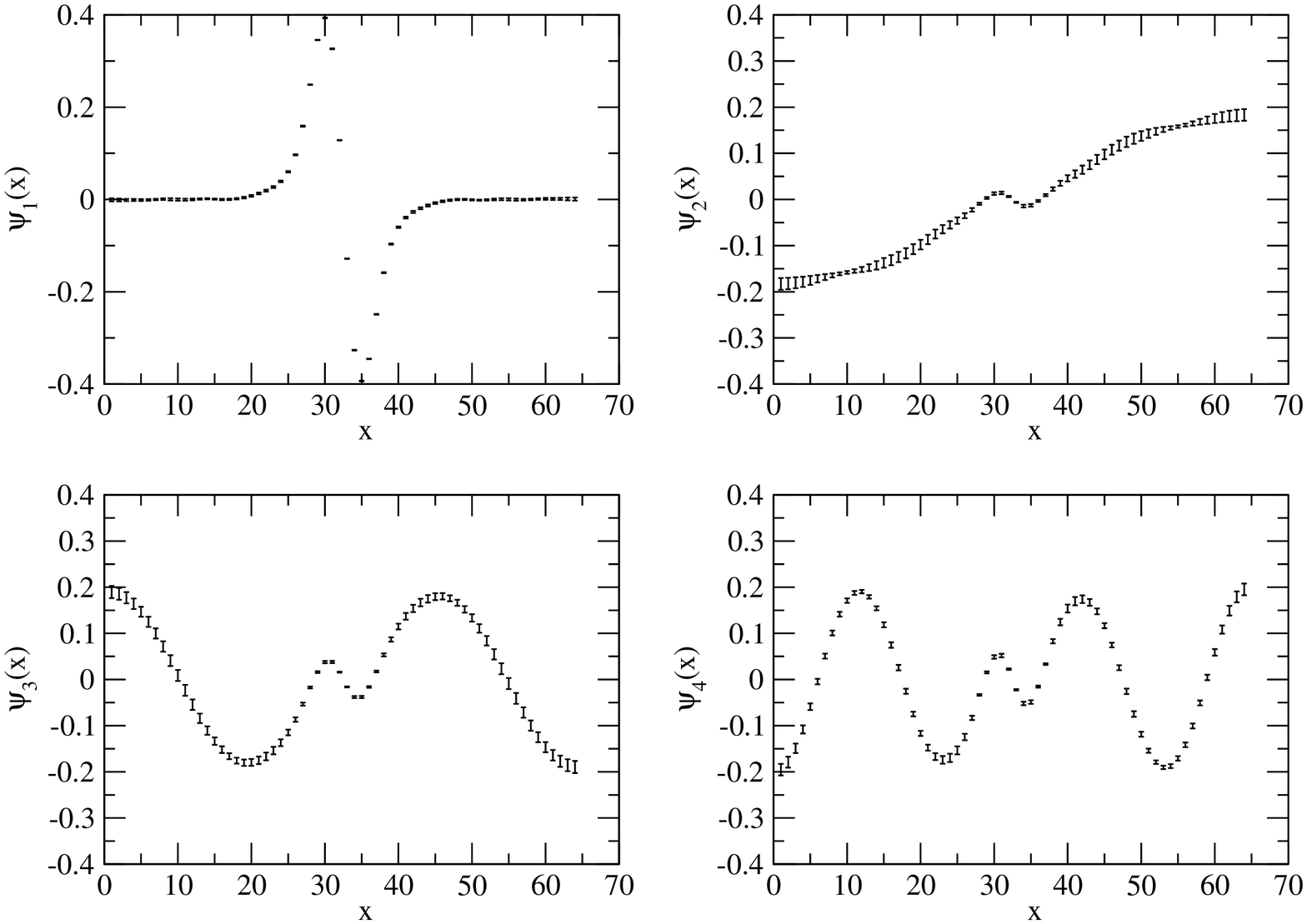}
\caption{\label{fig:l64wfs504} Plot of four lowest-lying wavefunctions for $L=64$, $m^2=0.504$.}
}

\section{Conclusions}

We have shown that correlation functions in the presence of a kink provide detailed information about the kink itself: its mass and its excitation spectrum, along with the approximate wavefunctions. One can measure these correlation functions using standard Monte Carlo techniques, thereby calculating its properties in a fully non-perturbative way in quantum field theory.

Quantum kink masses have been calculated previously in several different ways: at one-loop level in perturbation theory, and non-perturbatively using creation operators or by calculating the excess free energy of the kink across the two topological sectors. Our method, in which the mass is obtained from the kinetic energy of a kink with a known momentum, is closer in spirit to the way other masses are calculated in Monte Carlo simulations, and potential errors should be better controlled. 

Calculations of kink excitations have so far been restricted to the linear level, and it is quite likely that in many cases the interactions will modify even qualitative features of the spectrum, such as the number of bound states. It is therefore important to have a way of measuring the spectrum non-perturbatively. As a side product, we obtain approximate quantum wave functions of the energy eigenstates, which are only valid at weak coupling but which nevertheless help us to identify the states. 

We demonstrated our methods by carrying our Monte Carlo simulations at weak coupling in the $1+1$-dimensional $\lambda\phi^4$ theory. The results show that the mass and the excitations can be determined in practice, with relatively small errors. It will be interesting to extend these simulations to the critical region, and to see explicitly how the behaviour changes when critical phenomena become important and expectations based on linear theory become invalid. 

It should be straightforward to generalise our methods to other topological defects and more realistic theories, by considering correlators of suitable operators. Particularly interesting applications of this would be dyons as excitations of 't~Hooft-Polyakov monopoles, and localisation of degrees of freedom on cosmic strings, domain walls or solitonic branes in braneworld models. Comparing simulation results with known exact results would also provide a useful test of lattice formulations of supersymmetric field theories. 

\acknowledgments
We acknowledge support from the Science and Technology Facilities Council. This work has made use of the Imperial College \href{http://www3.imperial.ac.uk/ict/services/teachingandresearchservices/highperformancecomputing}{High Performance Computing Service}.

\bibliographystyle{JHEP}
\bibliography{kinkpaper}

\end{document}